\documentclass[preprint, onecolumn, tightenlines, superscriptaddress, nofootinbib, titlepage, floatfix]{revtex4-2}

\usepackage[T1]{fontenc}
\usepackage{amsmath,amssymb}
\usepackage{slashed}
\usepackage{bm}
\usepackage{graphicx}
\usepackage{lineno}
\usepackage{microtype}
\usepackage[
  colorlinks=true,
  linkcolor=blue,
  citecolor=blue,
  urlcolor=blue
]{hyperref}
\hypersetup{
  pdftitle={Chiral Response of QCD Topology across the Thermal Crossover},
  pdfauthor={Heng-Tong Ding}
}

\usepackage{placeins}

\newcommand{\chit}{\chi_{\rm top}}
\newcommand{\Ctop}{\mathcal C_{\rm top}}
\newcommand{\ml}{m_\ell}
\newcommand{\figw}{0.58\columnwidth}

\begin{document}
\title{Chiral Response of QCD Topology across the Thermal Crossover}

\author{Heng-Tong Ding}
\affiliation{Key Laboratory of Quark and Lepton Physics (MOE),
	Institute of Particle Physics, Central China Normal University,
	Wuhan 430079, China}

\author{Jin-Biao Gu}
\affiliation{INFN Sezione di Pisa,
	Largo Pontecorvo 3, I-56127 Pisa, Italy}

\author{Jos\'e Javier Hern\'andez Hern\'andez}
\affiliation{Key Laboratory of Quark and Lepton Physics (MOE),
	Institute of Particle Physics, Central China Normal University,
	Wuhan 430079, China}

\author{Sheng-Tai Li}
\affiliation{Key Laboratory of Quark and Lepton Physics (MOE),
	Institute of Particle Physics, Central China Normal University,
	Wuhan 430079, China}

\author{Jia Ni}
\affiliation{Key Laboratory of Quark and Lepton Physics (MOE),
	Institute of Particle Physics, Central China Normal University,
	Wuhan 430079, China}

\date{\today}

\begin{abstract}
	We introduce a novel relation connecting the light-quark-mass derivative
	of the topological susceptibility in quantum chromodynamics (QCD) to the
	connected correlation between the squared topological charge and the
	integrated light-quark scalar density. This relation gives direct access
	to the chiral response of QCD topology---the local logarithmic light-quark-mass derivative of the
	susceptibility---from a single ensemble at fixed quark masses for each
	temperature, without additional simulations at shifted masses.
	Applying this relation to $(2+1)$-flavor lattice QCD at
	$m_\pi\simeq220\,\mathrm{MeV}$ over temperatures ranging from well below
	to well above the chiral crossover, we find that the response rises
	rapidly through the crossover  from approximately unity
	toward two. This evolution corresponds to a change from nearly linear to nearly quadratic dependence of the topological susceptibility on the
	light-quark mass. Using the chiral response as the exponent of
a local power law in the light-quark mass, we determine its effective
amplitude from the topological susceptibility. This amplitude exhibits
a maximum near the crossover temperature, reflecting the competition
between the decreasing topological susceptibility and the increasing chiral response.
These results reveal a thermal reorganization of the
	quark-mass dependence of QCD topology beyond the suppression of
	topological fluctuations alone.
\end{abstract}

\maketitle

\section{Introduction}
Topological fluctuations of gluon fields generate the dependence of
the quantum chromodynamics (QCD) free energy on the strong-$CP$
angle $\theta$ \cite{Jackiw:1976pf}. They connect the chiral structure
of QCD \cite{tHooft:1976rip} to the strong-$CP$ problem
\cite{Peccei:1977hh} and the QCD axion
\cite{Weinberg:1977ma,Wilczek:1977pj}.
The topological susceptibility,
\begin{equation}
	\chit =
	\left.\frac{\partial^2 \mathcal{F}}{\partial\theta^2}\right|_{\theta=0}
	= \frac{\langle Q^2\rangle}{V_4},
	\label{eq:chi-def}
\end{equation}
measures the curvature of the free-energy density $\mathcal{F}$, where $Q$
is the total topological charge, $V_4$ is the Euclidean
four-volume, and $\langle\cdots\rangle$ denotes the QCD ensemble
average at $\theta=0$.
The susceptibility also determines the equilibrium QCD-axion mass
through $m_a^2(T)f_a^2=\chit(T)$
\cite{GrillidiCortona:2015jxo}.

Although $Q$ is defined by gluon fields, its equilibrium
fluctuations depend essentially on the quark masses.
The axial anomaly relates $\theta$ to the phase of the quark-mass
matrix, rendering $\theta$ unobservable if a quark is massless
\cite{tHooft:1976rip,GrillidiCortona:2015jxo}.
For two degenerate light quarks, $m_u=m_d=\ml$, the mass couples
to the scalar density $\bar u u+\bar d d$ and explicitly breaks
light-flavor chiral symmetry.
We define the \emph{chiral response of QCD topology} as the local
logarithmic derivative
$\Ctop=\partial\ln\chit/\partial\ln\ml$.
It measures the sensitivity of topological fluctuations to this
chiral symmetry-breaking source.
The thermal suppression of $\chit$ alone does not determine
whether its quark-mass dependence also changes.
Following $\Ctop$ through the crossover at fixed quark masses
therefore probes how the thermal medium responds to the same
mass source, and how the influence of spontaneous chiral symmetry
breaking on the $\theta$-dependent free energy evolves as the
chiral condensate melts.

The limiting regimes illustrate the physical content of this
chiral response of QCD topology. At low temperature, chiral effective theory gives an
approximately linear dependence $\chit\propto\ml$ for
$\ml\ll m_s$, with a leading coefficient related to the chiral
condensate \cite{Leutwyler:1992yt,GrillidiCortona:2015jxo}.
At sufficiently high temperature, the leading light-quark-mass
dependence $\chit\propto\ml^2$ at fixed $m_s$ is predicted by
both the dilute instanton gas approximation 
\cite{Gross:1980br,Schafer:1996wv} and the non-semiclassical
quasi-instanton description of chirally symmetric QCD~\cite{Kanazawa:2014cua}. Since $\Ctop$ measures the local mass exponent of $\chit$,
these predictions give $\Ctop\simeq 1$ at low temperature
and $\Ctop\simeq 2$ at high temperature.
Across these regimes, extensive lattice calculations have determined $\chit$ at zero
temperature \cite{Alles:2000cg,Bali:2001gk,Bonati:2015vqz},
through the crossover
\cite{Petreczky:2016vrs,Taniguchi:2016tjc,Gavai:2024mcj},
and at high temperatures
\cite{Borsanyi:2016ksw,Bonati:2018blm,Athenodorou:2022aay,
	Chen:2022fid,Bonanno:2024zyn,Kotov:2025ilm}.
Simulations at several sea-quark masses have explored its mass
dependence \cite{Bali:2001gk,Burger:2018fvb}, while mass derivatives
of relative topological-sector weights have been obtained from
scalar-condensate differences in fixed-topology simulations
\cite{Borsanyi:2016ksw}.

However, these results leave open how the local mass response
of the full susceptibility evolves through the strongly coupled
crossover. Inferring this chiral response $\Ctop$ from discrete mass points
requires finite differences, interpolation, or fitting.
A direct determination of $\Ctop$ would reveal where
and how rapidly the local mass dependence of topology departs
from the nearly linear behavior characteristic of the chirally
broken regime and approaches the leading quadratic behavior
expected at high temperature.

In this work, we establish a general relation expressing
quark-mass derivatives of the topological susceptibility through
connected correlations with scalar insertions, including
higher-order and mixed-flavor responses.
Its first-order form relates the mass derivative to the connected
correlation of $Q^2$ with the integrated light-quark scalar
density, providing a direct estimator on a single ensemble at
fixed quark masses for each temperature.
We apply this relation to $(2+1)$-flavor lattice QCD at
$m_\pi\simeq220\,\mathrm{MeV}$ in a fixed-scale temperature scan,
supplemented by a discretization check at a representative temperature
above the crossover.
The measured chiral response rises rapidly through the crossover from approximately unity toward two, revealing a
reorganization of the local light-quark-mass dependence of
QCD topology.

\section{Theoretical developments}
On a fixed gauge field, $Q$ has no explicit quark-mass dependence.
The mass response of $\chit$ therefore arises entirely from the
equilibrium weight. We derive the general response relations from
the continuum QCD partition function,
\begin{equation}
	\begin{aligned}
		\mathcal Z(\theta,\{m_f\})
		&=\int\mathcal DU\,e^{-S_g[U]+i\theta Q[U]}\\
		&\quad\times
		\exp\!\left[\sum_{f=u,d,s}W_f(m_f;U)\right],
	\end{aligned}
	\label{eq:mass-generator}
\end{equation}
where $S_g$ is the gauge action, $W_f(m_f;U)=\ln\det M_f[U]$ and
$M_f[U]= \slashed{D}[U]+m_f$, with $\slashed{D}[U]$ the massless
Euclidean Dirac operator in the gauge background $U$.
Thus $e^{W_f}=\det M_f$ is the weight obtained by integrating out
the quark field of flavor $f$.
The free-energy density is $\mathcal{F}=-V_4^{-1}\ln\mathcal Z$. 
For two degenerate light quarks, $m_u=m_d=\ml$, we vary both
light masses together at fixed $T$, $V_4$, $m_s$, and gauge
action. With $W_\ell=W_u+W_d=2\ln\det M_\ell$, we define
the insertions
\begin{equation}
	X_k(U)\equiv
	\frac{\partial^k W_\ell(\ml;U)}{\partial\ml^k},
	\qquad k\geq1.
	\label{eq:mass-insertions}
\end{equation}
Since $M_\ell=\slashed{D}+m_\ell$ is linear in the common light-quark
mass, $X_k=2(-1)^{k-1}(k-1)!\operatorname{Tr}M_\ell^{-k}$.
The trace includes integration over spacetime and sums over color
and Dirac indices. In particular, $X_1=2\operatorname{Tr}M_\ell^{-1}$
is the integrated light-quark scalar insertion. Equivalently,
differentiating the Euclidean quark integral at fixed $U$ identifies
$X_1$ with the fermionic expectation value of
$-\int d^4x\,(\bar u u+\bar d d)$.

Repeated mass differentiation of the normalized measure generates
joint cumulants involving $Q^2$ and the scalar insertions.
To organize the contributions at order $n\geq1$,  let $\mathcal P_n$ contain the nonnegative integer vectors
$\bm p=(p_1,\ldots,p_n)$ satisfying $\sum_{k=1}^n kp_k=n$.
The subscript $c$ below denotes a joint cumulant, with $Q^2$ counted as
one entry and $X_k$ repeated $p_k$ times.
At $\theta=0$, $CP$ symmetry makes correlators with an odd number of
$Q$ insertions vanish. Differentiating the generating function then gives
\begin{equation}
	\begin{aligned}
		\frac{\partial^n\chit}{\partial\ml^n}
		&=-\frac{1}{V_4}
		\left.
		\frac{\partial^{n+2}\ln\mathcal Z}
		{\partial\theta^2\partial\ml^n}
		\right|_{\theta=0}\\
		&=\frac{1}{V_4}
		\sum_{\bm p\in\mathcal P_n}
		\frac{
			n!\left\langle Q^2\prod_{k=1}^nX_k^{p_k}\right\rangle_c
		}{
			\prod_{k=1}^n p_k!(k!)^{p_k}
		}.
	\end{aligned}
	\label{eq:nth-response}
\end{equation}
Equation~\eqref{eq:nth-response} connects arbitrary-order mass
derivatives of the topological susceptibility to correlations
between topological fluctuations and scalar insertions.
Keeping the flavor masses independent generates mixed-flavor
derivatives in the same way.

For the numerical study below, we focus on the first-order
response.  Setting $n=1$ in
Eq.~\eqref{eq:nth-response} gives
\begin{equation}
	\frac{\partial\chit}{\partial\ml}
	=\frac{1}{V_4}\left\langle Q^2X_1\right\rangle_c,
	\label{eq:FH}
\end{equation}
and its normalized response to the common light-quark mass,
\begin{equation}
	\Ctop(T)\equiv
	\frac{\partial\ln\chit}{\partial\ln\ml}
	=
	\ml\,
	\frac{\left\langle Q^2X_1\right\rangle_c}
	{\langle Q^2\rangle}.
	\label{eq:Ctop}
\end{equation}

Equation~\eqref{eq:Ctop} measures the local logarithmic mass
slope, reducing to $\Ctop=p$ for $\chit\propto\ml^p$.
It is invariant under mass-independent multiplicative
renormalization of the quark masses.
Here the joint cumulant is the covariance,
$\langle Q^2X_1\rangle_c
=\langle Q^2X_1\rangle-\langle Q^2\rangle\langle X_1\rangle$,
so configuration-independent additive contributions to $X_1$
cancel. 
Since $Q^2$ and $X_1$ are evaluated on the same configurations
at the simulated quark masses, Eqs.~\eqref{eq:FH}
and~\eqref{eq:Ctop} give the local mass response directly
from equilibrium fluctuations on a single ensemble.

\section{Lattice QCD calculations}
To follow the chiral response across the thermal crossover, we evaluate
Eq.~\eqref{eq:Ctop} on $(2+1)$-flavor ensembles generated with
the highly improved staggered quark action
\cite{Follana:2006rc} and a tree-level Symanzik-improved gauge action~\cite{Weisz:1982zw}.
Our temperature scan follows the fixed-scale lattice setup of
Ref.~\cite{Ding:2021cwv}, with $a\simeq0.117\,\mathrm{fm}$.
The strange-quark mass is physical, while the
light-quark masses are heavier than physical, with $\ml=m_s/10$,
corresponding to $m_\pi=220.61(6)\,\mathrm{MeV}$; the spatial
lattice size is $32^3$ with $m_\pi L\simeq4.2$.
Varying $N_\tau$ at fixed quark masses gives $T=(aN_\tau)^{-1}$
from $70.2$ to $280.9\,\mathrm{MeV}$. The chiral crossover temperature for these ensembles is $T_{\rm pc}\simeq170\,\mathrm{MeV}$~\cite{Ding:2021cwv}.

To assess discretization effects, we choose a representative temperature
above the crossover, $T=210.8\,\mathrm{MeV}$, and compare the original
$32^3\times8$ ensemble with additional $40^3\times10$ and
$48^3\times12$ ensembles.
The quark masses are matched across the three ensembles, with physical
$m_s$ and $m_\pi\simeq220\,\mathrm{MeV}$.
At this temperature, their common
aspect ratio $N_\sigma/N_\tau=4$ also fixes the physical spatial
volume. Table~\ref{tab:ensembles} in Appendix~\ref{app:ensembles}
lists the lattice parameters for all ten ensembles, with about 4000--15000 saved configurations per ensemble.

We compute $Q^{\rm lat}$ after Wilson flow~\cite{Luscher:2010iy} from an
improved topological charge density that reduces the
tree-level discretization error from $\mathcal{O}(a^2)$ to
$\mathcal{O}(a^4)$~\cite{Bilson-Thompson:2002xlt}, using the \texttt{SIMULATeQCD} software
suite~\cite{Mazur:2021zgi,HotQCD:2023ghu}. We choose a flow time $\tau_f T^2=1/10$,
which lies on the observed flow-time plateaus and satisfies
$\tau_f T^2<1/8$~\cite{Petreczky:2016vrs}. Rounding the flowed charge
to the nearest integer gives the sector label
$Q=\operatorname{round}(Q^{\rm lat})$ used for eigenvalue reweighting.
Appendix~\ref{app:flow_sampling} shows the flow-time dependence and
the sampling of topological sectors. 

For the scalar insertion, we write the light-quark matrix in lattice units as
$\widehat M_\ell=\widehat{\slashed{D}}+\widehat m_\ell$, with
$\widehat m_\ell=a\ml$.
The four-taste staggered determinant enters
with a power of $1/4$ per flavor, so the two degenerate light quarks
carry the combined weight $[\det\widehat M_\ell]^{1/2}$. Differentiating its logarithm with respect to their common
mass $\widehat m_\ell$ gives
$\widehat X_1=\tfrac12\operatorname{Tr}\widehat M_\ell^{-1}$.
Using this insertion and $\widehat m_\ell$ in Eq.~\eqref{eq:Ctop} gives
the chiral response of QCD topology in the simulated measure.

At nonzero lattice spacing, the would-be topological zero modes of the
staggered operator are displaced from zero. 
To restore their continuum
mass factors, we set the selected eigenvalues to zero in the rooted
sea-quark determinant~\cite{Borsanyi:2016ksw,Kovacs:2017pox,Brandt:2024gso}.
The ratio of the modified determinant to the original one on each
configuration $U$ defines the reweighting factor $w(U)$.
With $\langle O\rangle_{\rm rw}=\langle wO\rangle/\langle w\rangle$,
the reweighted susceptibility is $\chit^{\rm rw}=\langle Q^2\rangle_{\rm rw}/V_4$.
Since both the rooted determinant and $w$ depend on the common
light-quark mass, Eq.~\eqref{eq:Ctop} becomes
\begin{equation}
 \begin{aligned}
 \Ctop^{\rm rw}
 &=\widehat m_\ell\,
 \frac{\langle Q^2\widehat X_1^{\rm rw}\rangle_{{\rm rw},c}}
 {\langle Q^2\rangle_{\rm rw}},\\
 \widehat X_1^{\rm rw}
 &=\frac12\operatorname{Tr}\widehat M_\ell^{-1}
   +\frac{\partial\ln w}{\partial\widehat m_\ell}.
 \end{aligned}
 \label{eq:staggered-response}
\end{equation}
We evaluate Eq.~\eqref{eq:staggered-response} at the simulated masses
using $Q^2$, a stochastic estimate of the scalar trace, and the selected
low eigenvalues, all measured on the same configurations.
Appendix~\ref{app:lat_ctop} gives the eigenvalue expressions
for $w$ and its light-quark mass derivative and specifies the
normalization of the measured trace. All results below use the reweighted measure, and
we omit the superscript `$\mathrm{rw}$'.

\section{Results}
\begin{figure}[!h]
	\includegraphics[width=\figw]{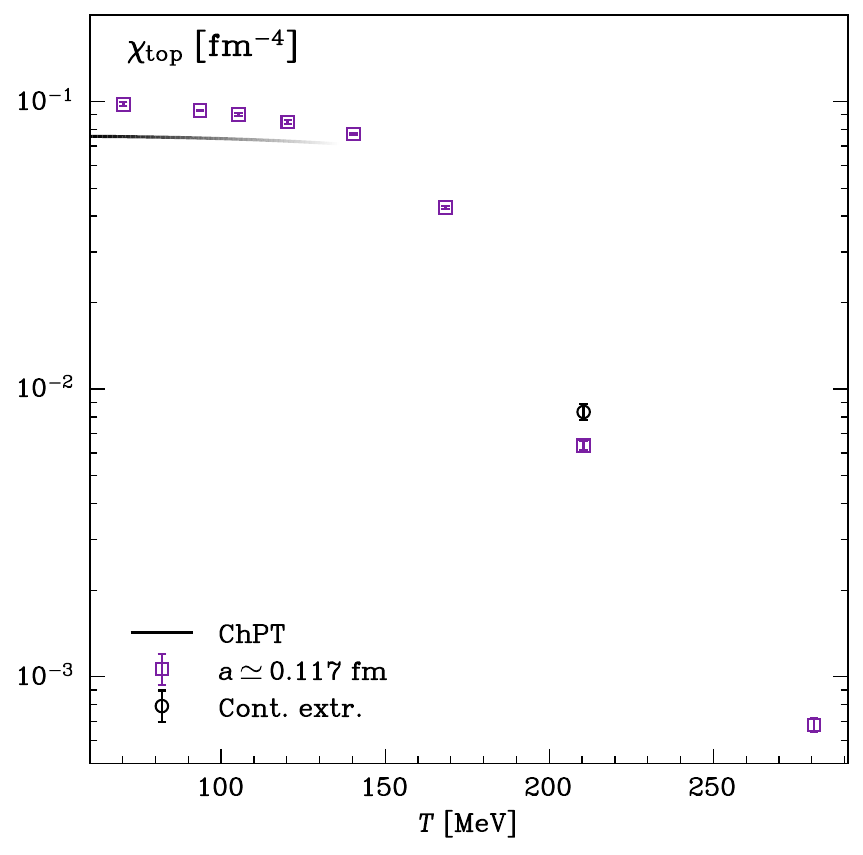}
	\caption{$\chit$ as a function of temperature. Purple squares denote results at $a\simeq0.117\,\mathrm{fm}$, while the
		black circle shows the continuum extrapolation at $T\simeq211\,\mathrm{MeV}$
		using $N_\tau=8,10,12$ ensembles at matched masses and physical volume.
		The solid curve represents the low-temperature ChPT reference~\cite{GrillidiCortona:2015jxo}.}
	\label{fig:chi}
\end{figure}

Fig.~\ref{fig:chi} shows the temperature dependence of the
topological susceptibility $\chit$ at $a\simeq0.117\,\mathrm{fm}$.
The susceptibility decreases from about
$9.8\times10^{-2}$ to $6.8\times10^{-4}\, \mathrm{fm}^{-4}$
over $T\approx70$--$281\,\mathrm{MeV}$, corresponding to a suppression
of roughly two orders of magnitude. This thermal suppression is
qualitatively consistent with earlier results
\cite{Bonati:2015vqz,Petreczky:2016vrs,Gavai:2024mcj}.
The lowest-temperature results lie above the chiral perturbation
theory (ChPT) reference \cite{GrillidiCortona:2015jxo}.

\begin{figure}[!h]
	\includegraphics[width=\figw]{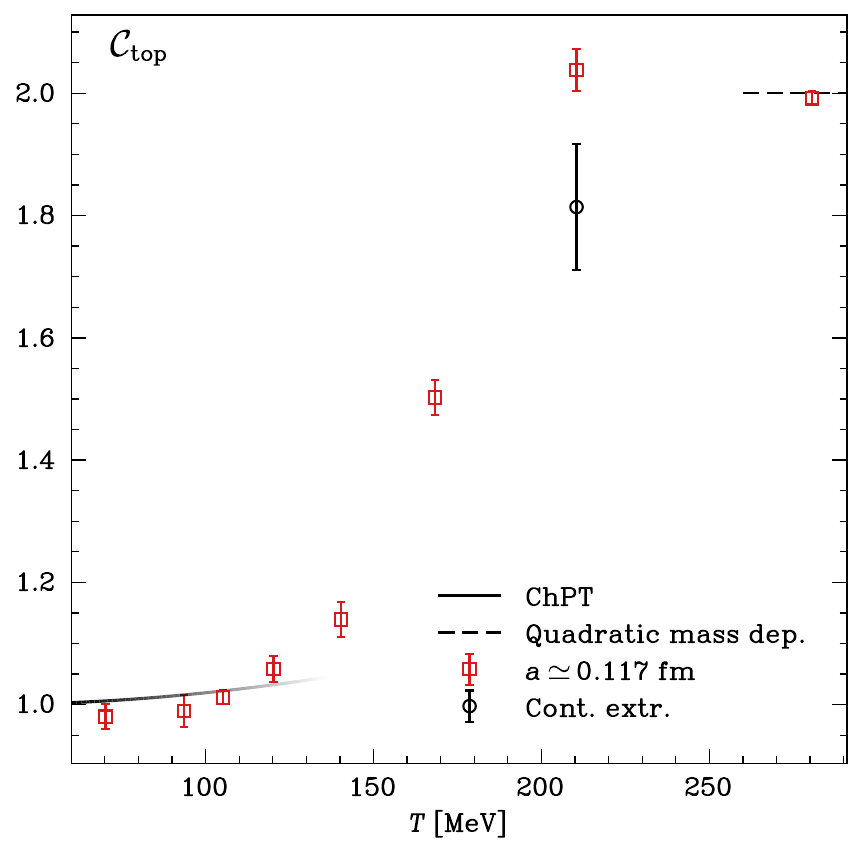}
	\caption{Temperature dependence of the chiral response of QCD topology
		$\Ctop=\partial\ln\chit/\partial\ln\ml$. Red squares denote results at $a\simeq0.117\,\mathrm{fm}$, while the
		black circle shows the continuum extrapolation at $T\simeq211\,\mathrm{MeV}$. 
		The solid curve represents the low-temperature ChPT reference \cite{GrillidiCortona:2015jxo}, while
		the dashed line denotes $\Ctop=2$, corresponding to the leading
		quadratic light-quark-mass dependence expected at high temperature~\cite{Gross:1980br,Schafer:1996wv,Kanazawa:2014cua}. 
        }
	\label{fig:response}
\end{figure}

The accompanying change in the local mass dependence is revealed by
$\Ctop$ in Fig.~\ref{fig:response}. At leading order in ChPT at zero temperature, $\chi_{\rm top}^{\rm ChPT} \simeq \Sigma\,\ml/2$,
with $\Sigma$ the chiral condensate in the chiral limit~\cite{Leutwyler:1992yt,GrillidiCortona:2015jxo},
which gives $\Ctop^{\rm ChPT} \simeq 1$. The response rises from near the
ChPT value at low temperature to about $1.5$ at
$T=168.5\,\mathrm{MeV}$ and nearly two at $210.8$ and
$280.9\,\mathrm{MeV}$. Thus, the local light-quark-mass dependence
changes from nearly linear toward nearly quadratic through the
crossover. A factorized form $\chit(T,\ml)=F(T)g(\ml)$ would give
a temperature-independent $\Ctop$; the observed rise therefore
reveals a change in the mass dependence beyond thermal suppression.

We characterize this mass dependence at each temperature through the
local power-law representation
$\chit(T,H)\simeq A_{\mathrm{eff}}(T)H^{p(T)}$, where $H=\ml/m_s$.
Matching $\chit$ and its logarithmic mass derivative at the
simulated mass ratio $H=1/10$ gives $p(T)=\Ctop(T)$ and the
effective local power-law amplitude
$A_{\mathrm{eff}}(T)=\chit(T)H^{-\Ctop(T)}$.

\begin{figure}[!h]
	\includegraphics[width=\figw]{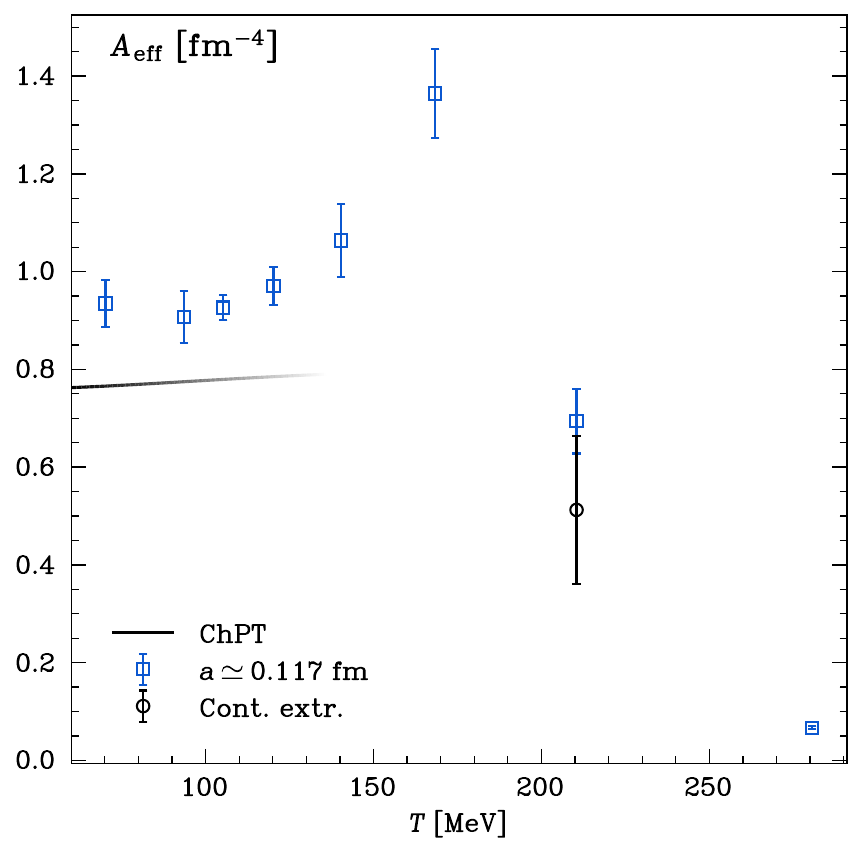}
	\caption{Temperature dependence of the effective local power-law amplitude $A_{\mathrm{eff}}=\chi_{\mathrm{top}}H^{-\Ctop}$ at $H=1/10$. Blue squares denote results at $a\simeq0.117\,\mathrm{fm}$, while  the black circle represents the continuum extrapolation at $T\simeq211\,\mathrm{MeV}$.
		The solid curve shows the ChPT reference \cite{GrillidiCortona:2015jxo}.}
	\label{fig:amplitude}
\end{figure}

Fig.~\ref{fig:amplitude} shows that $A_{\mathrm{eff}}$ rises from about
$0.93\,\mathrm{fm}^{-4}$ at low temperature to its largest
sampled value, about $1.36\,\mathrm{fm}^{-4}$, at
$T=168.5\,\mathrm{MeV}$, near $T_{\rm pc}\simeq170\,\mathrm{MeV}$.
It then decreases to about $0.067\,\mathrm{fm}^{-4}$ at
$280.9\,\mathrm{MeV}$.
At fixed $H$, the increasing chiral response $\Ctop$ enhances $H^{-\Ctop}$ while
$\chit$ decreases. Their competition produces the observed maximum,
with thermal suppression dominating above the crossover.

At the two highest temperatures of the fixed-spacing scan,
$T=210.8$ and $280.9\,\mathrm{MeV}$, where $\Ctop$ is close to two,
we also consider the quadratic--quartic ansatz
$\chit(T,H)\simeq A_2(T)H^2+A_4(T)H^4$.
Matching $\chit$ and its mass derivative gives
$A_2=(4-\Ctop)\chit/(2H^2)$ and
$A_4=(\Ctop-2)\chit/(2H^4)$.
We find the quadratic coefficient $A_2$ close to the effective
amplitude $A_{\mathrm{eff}}$ at these two temperatures,
as expected from
$A_2/A_{\mathrm{eff}}=\tfrac12(4-\Ctop)H^{\Ctop-2}\to1$ as
$\Ctop\to2$.
In particular, 
we find $A_2/A_{\mathrm{eff}} = 0.90(9)$ and $1.02(3)$ at $T = 210.8$ and $280.9\,\mathrm{MeV}$, respectively. The corresponding quartic
corrections relative to the quadratic term, $A_4H^2/A_2$, are
$(1.9\pm1.8)\%$ and $(-0.4\pm0.6)\%$.

To assess discretization effects, we compare these results with the
continuum estimates at $T\simeq211\,\mathrm{MeV}$, obtained from
matched $N_\tau=8,10,12$ ensembles and shown as black circles in
Figs.~\ref{fig:chi}--\ref{fig:amplitude}.
The topological susceptibility approaches the continuum from below, yielding
$\chit=0.00833(54)\,\mathrm{fm}^{-4}$ in our setup with $m_\pi\simeq220\,\mathrm{MeV}$.
The continuum extrapolation lowers $\Ctop$ to $\Ctop = 1.81(10)$, which remains appreciably above
the near-linear benchmark at low temperature.
The corresponding continuum-extrapolated $A_{\mathrm{eff}} = 0.51(15)\,\mathrm{fm}^{-4}$ lies below the fixed-spacing value and is compatible within uncertainties.
For comparison with existing physical-mass results, 
we assume that $\Ctop$ retains its measured continuum value between $H=1/10$ and $H_{\rm phys}$, where
$H_{\rm phys}^{-1}=27.42(12)$~\cite{FlavourLatticeAveragingGroupFLAG:2024oxs}. Rescaling the continuum susceptibility as $\chit\propto H^{\Ctop}$
then gives $\chit^{\rm phys}=0.00134(15)\,\mathrm{fm}^{-4}$, in agreement with
$0.00133(14)\,\mathrm{fm}^{-4}$ obtained from
Ref.~\cite{Borsanyi:2016ksw} at the same temperature and with the same
light-quark mass convention.

\section{Conclusions}
In this work, we have established relations connecting
quark-mass derivatives of topological observables to connected
correlations with scalar insertions. These relations give direct access to
$\partial^n\chit/\partial\ml^n$ for any $n\geq 1$ on a single
ensemble at each temperature, without additional simulations at
shifted light-quark masses. 

Applying the first-order relation to $(2+1)$-flavor lattice QCD at
$m_\pi\simeq220\,\mathrm{MeV}$ and $a\simeq0.117\,\mathrm{fm}$,
we find that the chiral response of QCD topology, $\Ctop$, rises
from approximately one toward two through the crossover, 
while $\chit$ falls by about two orders of magnitude through the crossover.  At $T\simeq211\,\mathrm{MeV}$, the continuum-extrapolated
response, $\Ctop=1.81(10)$, remains well above the low-temperature
benchmark of unity. In our temperature scan, the effective local power-law amplitude $A_{\mathrm{eff}}$ peaks near $T_{\rm pc}$, reflecting the
competition between the thermal suppression of $\chit$ and the
rise of $\Ctop$.
Together, these results reveal a thermal reorganization of
the quark-mass dependence of topology beyond the suppression
of topological fluctuations alone.

Calculations at finer lattice spacings will extend the
continuum analysis across the crossover.
The formalism can also be implemented with other fermion
discretizations, enabling independent tests of these findings.
Extending the analysis to lighter quark masses will directly
test the mass scaling of $\chit$ toward the chiral limit,
providing a topological probe of effective $U(1)_A$
symmetry restoration \cite{Ding:2026blh}.
The same construction gives mass responses of higher
topological-charge cumulants and of coefficients such as
$b_2$ in the $\theta$ expansion of the free energy.
It thus extends the analysis to the quark-mass dependence
of non-Gaussian topological fluctuations and of the axion
potential beyond its curvature, including in a
background magnetic field.

\section*{Acknowledgements}
This work is supported partly by the National Natural Science Foundation of China under Grants  No. 12325508, No. 12293064 and No. 12293060, as well as the National Key Research and Development Program of China under Contract No. 2022YFA1604900 and the Fundamental Research Funds for the Central Universities, Central China Normal University under Grants No. 30101250314 and No.30106250152. The numerical simulations have been performed on the GPU cluster in the Nuclear Science Computing Center at Central China Normal University ($\mathrm{NSC}^{3}$) and Wuhan Supercomputing Center.

\bibliographystyle{JHEP}
\bibliography{references}

\setcounter{secnumdepth}{1}
\appendix
\setcounter{equation}{0}
\renewcommand{\theequation}{\thesection\arabic{equation}}
\section{Lattice ensembles}
\label{app:ensembles}
The fixed-scale scan comprises eight $32^3\times N_\tau$ ensembles at
$a\simeq0.117\,\mathrm{fm}$ and fixed quark masses,
with $4\,188$ to $15\,114$ gauge configurations per temperature
(Table~\ref{tab:ensembles}). For the discretization check at
$T=210.8\,\mathrm{MeV}$, we supplement the $32^3\times8$ ensemble
with $40^3\times10$ and $48^3\times12$ ensembles at matched quark
masses. All three have $N_\sigma/N_\tau=4$, with
$L=aN_\sigma \simeq3.74\,\mathrm{fm}$ and $m_\pi L\simeq4.2$.
On each configuration, we estimate the light-quark trace
$\operatorname{Tr}\widehat M_\ell^{-1}$ using 102 random source vectors for ensembles at $a\simeq0.117\,\mathrm{fm}$ and 55 random source vectors for the rest.

\begin{table}[ht]
 \begin{ruledtabular}
 \begin{tabular}{cccc}
 $N_\sigma^3\times N_\tau$ & $a$ [fm] & $T$ [MeV] & $N_{\rm conf}$ \\
 \hline
 $32^3\times24$ & 0.117  &  70.2 &  6517 \\
 $32^3\times18$ & 0.117  &  93.6 & 15114 \\
 $32^3\times16$ & 0.117  & 105.3 & 13574 \\
 $32^3\times14$ & 0.117  & 120.4 &  6980 \\
 $32^3\times12$ & 0.117  & 140.4 &  8001 \\
 $32^3\times10$ & 0.117  & 168.5 &  8867 \\
 $32^3\times8$  & 0.117  & 210.8 &  5698 \\
 $32^3\times6$  & 0.117  & 280.9 &  4188 \\
 \hline
 $40^3\times10$ & 0.0936 & 210.8 &  4382 \\
 $48^3\times12$ & 0.0780 & 210.8 &  6209 \\
 \end{tabular}
 \end{ruledtabular}
  \caption{Lattice parameters and gauge-configuration counts.
 All ensembles have a physical strange-quark mass and $\ml=m_s/10$,
 corresponding to $m_\pi\simeq220\,\mathrm{MeV}$. The first eight rows give our temperature scan using the fixed-scale
lattice setup of Ref.~\cite{Ding:2021cwv}.}
   \label{tab:ensembles}
\end{table}

\section{Flow-time dependence, topological sampling, and continuum extrapolation}
\label{app:flow_sampling}

\begin{figure}[p]
	\centering
	\includegraphics[width=0.62\columnwidth]{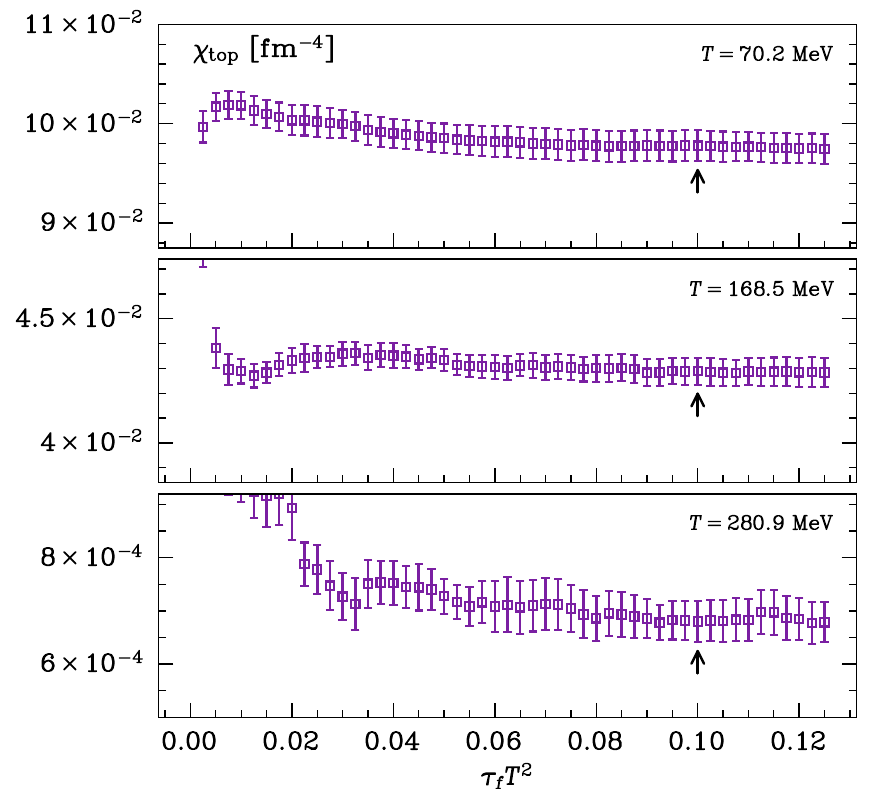}
	\caption{Topological susceptibility $\chit$ as a function of the
		dimensionless flow time $\tau_f T^2$ for $T=70.2$, $168.5$ and
		$280.9$~MeV (top to bottom), obtained on $32^3\times N_\tau$ lattices
		with $N_\tau=24$, $10$ and $6$ at $a\simeq0.117$~fm. The arrows
		indicate $\tau_f T^2=0.1$, the flow time used in the analysis.}
	\label{fig:chitop_flow}

	\vspace{1.5em}

	\includegraphics[width=\figw]{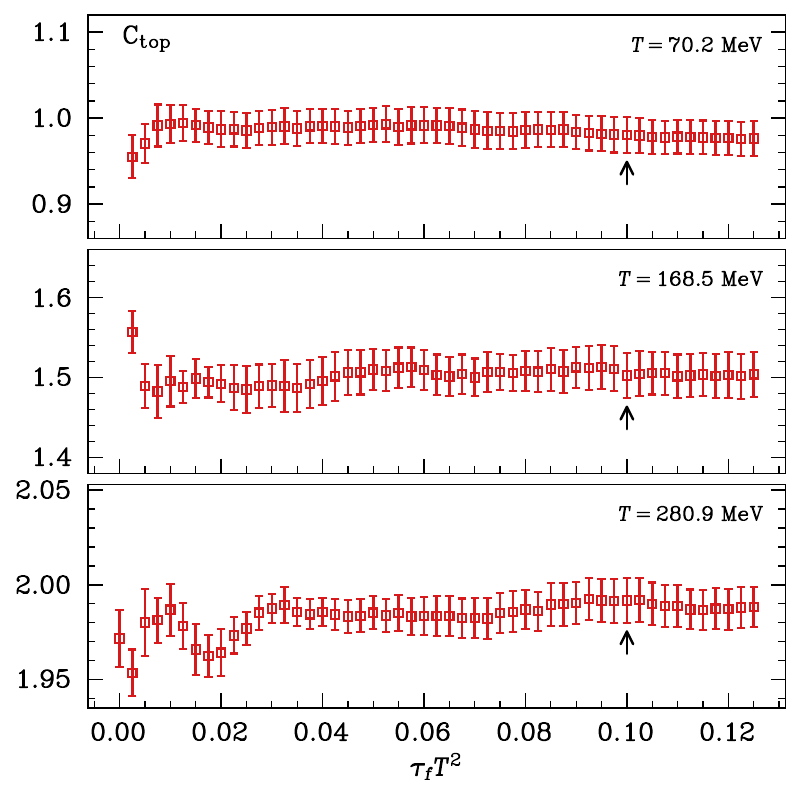}
	\caption{Chiral response $\Ctop$ as a function of the dimensionless
		flow time $\tau_f T^2$ for the same temperatures and panel ordering
		as in Fig.~\ref{fig:chitop_flow}. The arrows indicate the flow time
		used in the analysis, $\tau_f T^2=0.1$.}
	\label{fig:ctop_flow}
\end{figure}

Figures~\ref{fig:chitop_flow} and~\ref{fig:ctop_flow} show $\chit$ and
$\Ctop$, respectively, as functions of the flow time $\tau_f T^2$ at three
representative temperatures. As the Wilson flow smooths the gauge fields,
$Q^{\rm lat}$ approaches integer values. At each temperature, both
observables reach plateaus that include our reference flow time
$\tau_f T^2=0.1$. We integrate the flow equation using the adaptive
third-order Runge--Kutta method implemented in the \texttt{SIMULATeQCD}
software suite~\cite{Mazur:2021zgi,HotQCD:2023ghu}.

\begin{figure}[htbp]
	\centering
	\includegraphics[width=\figw]{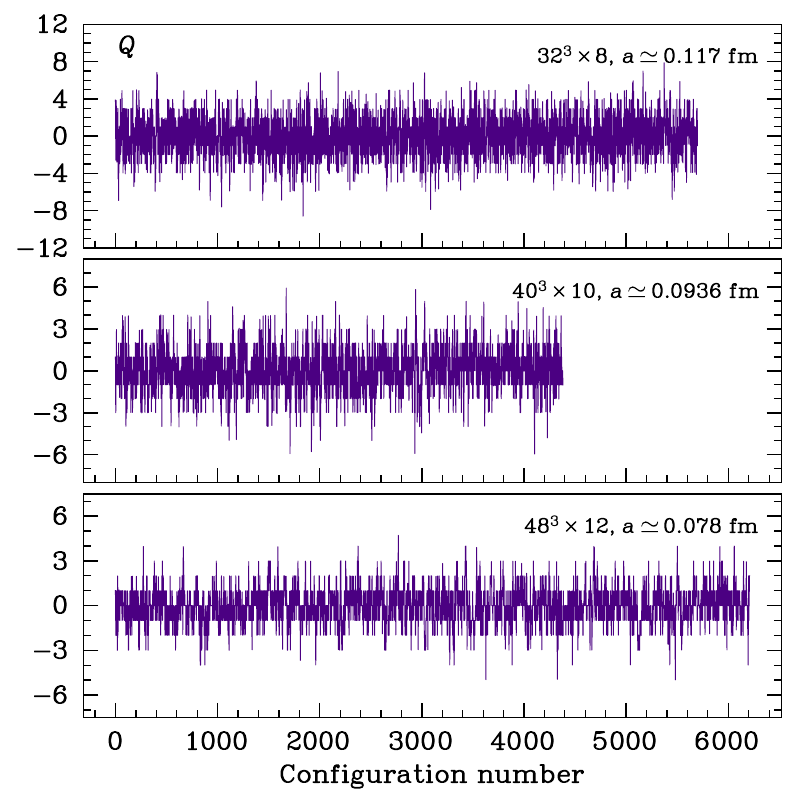}
	\caption{Monte Carlo histories of the unreweighted topological charge
		$Q$ on the $32^3\times8$, $40^3\times10$, and $48^3\times12$
		ensembles (top to bottom), with lattice spacings $a\simeq0.117$,
		$0.0936$, and $0.0780$~fm, respectively.}
	\label{fig:Qtop_history}
\end{figure}

Figure~\ref{fig:Qtop_history} shows the Monte Carlo histories of the
unreweighted topological charge $Q$ on the three ensembles at
$T\simeq211$~MeV used for the discretization check. All three histories
show frequent transitions between topological sectors, with no sign of
topological freezing. Since the reweighting suppresses the $Q\neq0$
sectors, the reweighted $\langle Q^2\rangle_{\rm rw}$ is smaller than
suggested by these histories.

\begin{figure}[htbp]
    \centering
	\includegraphics[width=\figw]{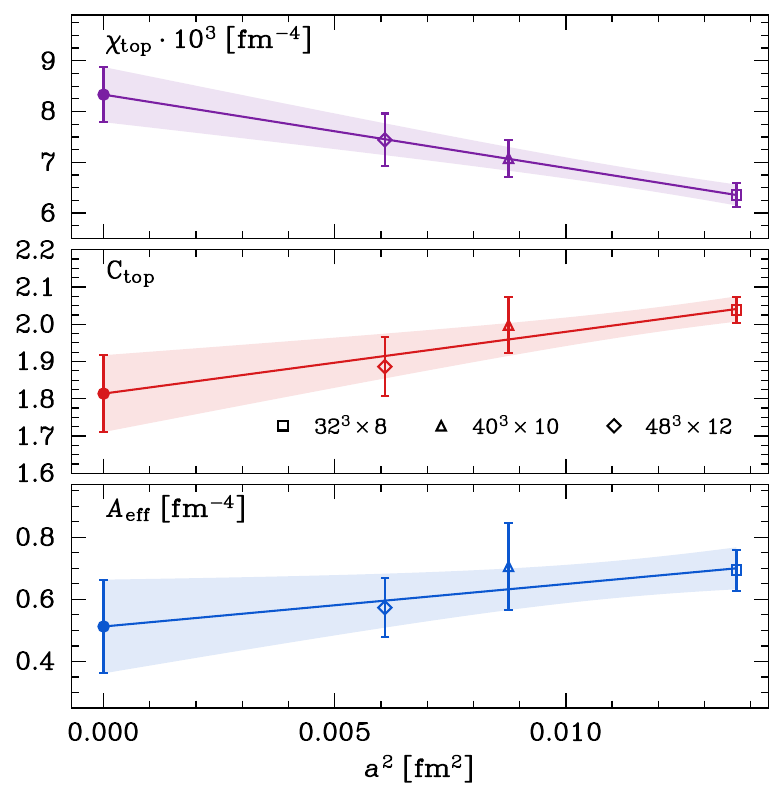}
	\caption{Continuum extrapolations at $T\simeq211$~MeV of $\chit$ (top),
		$\Ctop$ (middle), and $A_{\rm eff}$ (bottom), all evaluated at
		$\tau_f T^2=0.1$. Open squares, triangles, and diamonds denote the
		$32^3\times8$, $40^3\times10$, and $48^3\times12$ results,
		respectively. Filled circles are the continuum values. Lines and
		bands show linear fits in $a^2$ and their jackknife uncertainties.}
	\label{fig:clim_T211}
\end{figure}

Using these three ensembles, we extrapolate each observable to the
continuum with a linear fit in $a^2$ on each jackknife sample
(Fig.~\ref{fig:clim_T211}). The reweighted susceptibility $\chit$
increases toward the continuum, as also seen at high temperature in
Ref.~\cite{Borsanyi:2016ksw}. On the coarsest lattice, $\chit$ lies
about $24\%$ below its continuum estimate. The fractional cutoff effect
is smaller for $\Ctop$, whose coarsest-lattice value lies about $12\%$
above its continuum estimate of $1.81(10)$. The central value of
$A_{\rm eff}$ likewise decreases toward the continuum, although all
three finite-spacing results agree with the extrapolated value within
uncertainties.

\section{Lattice determination of $\Ctop$}
\label{app:lat_ctop}

Eigenvalue reweighting has been tested in calculations of the
topological susceptibility at high temperatures~\cite{Borsanyi:2016ksw,Kovacs:2017pox}
and near the crossover~\cite{Brandt:2024gso}. We hereby summarize its
implementation for $\Ctop$. 

On each configuration $U$, we measure
$Q^{\rm lat}$ after Wilson flow~\cite{Luscher:2010iy} at
$\tau_f T^2=1/10$ and assign the integer sector
$\nu(U)=Q(U)=\operatorname{round}(Q^{\rm lat}(U))$. In the continuum,
the topological zero modes in sector $\nu$ contribute a factor
$m_f^{|\nu|}$ to the determinant for each flavor $f$. At finite lattice
spacing, the corresponding staggered modes are displaced from zero.
We denote the $2|\nu|$ selected eigenvalue pairs of the massless
sea-quark operator for flavor $f$ by $\pm i\widehat\lambda_{j,f}$,
with $\widehat\lambda_{j,f}>0$.

In lattice units, $\widehat M_f=\widehat{\slashed{D}}+\widehat m_f$, with
$\widehat m_f=am_f$. For two degenerate light quarks and one strange
quark, the rooted fermion weight is
$[\det\widehat M_\ell]^{1/2}[\det\widehat M_s]^{1/4}$. 
For each selected eigenvalue pair, we replace the factor
$\widehat\lambda_{j,f}^{\,2}+\widehat m_f^2$ in $\det\widehat M_f$
by its zero-mode value $\widehat m_f^2$. The ratio of the modified
rooted weight to the original rooted weight is
\begin{equation}
 w(U)=\prod_{j=1}^{2|\nu|}
 \left(\frac{\widehat m_\ell^2}
 {\widehat\lambda_{j,\ell}^{\,2}+\widehat m_\ell^2}\right)^{1/2}
 \left(\frac{\widehat m_s^2}
 {\widehat\lambda_{j,s}^{\,2}+\widehat m_s^2}\right)^{1/4}.
 \label{eq:app-reweighting-factor}
\end{equation}
For $\nu=0$, the product is empty and $w=1$. With $\langle\cdot\rangle$
denoting the simulated ensemble average, the reweighted expectation
value and the susceptibility at $\theta=0$ are
\begin{equation}
 \langle O\rangle_{\rm rw}=\frac{\langle wO\rangle}{\langle w\rangle},
 \qquad
 \chit^{\rm rw}=\frac{\langle Q^2\rangle_{\rm rw}}{a^4N_4},
 \label{eq:app-reweighted-expectation}
\end{equation}
where $N_4=N_\sigma^3N_\tau$ is the number of lattice sites.

At fixed $U$, $Q^2$ has no explicit light-quark mass dependence.
Differentiating the normalized measure with respect to the common
mass $\widehat m_\ell$ therefore gives
\begin{equation}
 \begin{aligned}
 \frac{\partial\langle Q^2\rangle_{\rm rw}}{\partial\widehat m_\ell}
 &=\langle Q^2\widehat X_1^{\rm rw}\rangle_{{\rm rw},c},\\
 \widehat X_1^{\rm rw}
 &=\frac12\operatorname{Tr}\widehat M_\ell^{-1}
   +\frac{\partial\ln w}{\partial\widehat m_\ell},
 \end{aligned}
 \label{eq:app-reweighted-covariance}
\end{equation}
where $c$ denotes the covariance in the reweighted ensemble. The trace
term comes from the rooted light-quark determinant. Holding the sector,
selected massless eigenvalues, and strange-quark mass fixed gives
\begin{equation}
 \frac{\partial\ln w}{\partial\widehat m_\ell}
 =\sum_{j=1}^{2|\nu|}
 \left(\frac{1}{\widehat m_\ell}
 -\frac{\widehat m_\ell}
 {\widehat\lambda_{j,\ell}^{\,2}+\widehat m_\ell^2}\right).
 \label{eq:app-weight-derivative}
\end{equation}
This correction replaces the selected pairs' contributions to the
scalar insertion by their zero-mode values, as in the 
estimators of chiral condensates in Refs.~\cite{Borsanyi:2016ksw,Brandt:2024gso}.

On each configuration, we estimate $\operatorname{Tr}\widehat M_\ell^{-1}$
stochastically. For the trace per color
and lattice site, $d_\ell=\operatorname{Tr}\widehat M_\ell^{-1}/(3N_4)$,
the corrected density is
$d_\ell^{\rm rw}=d_\ell+[2/(3N_4)]\,\partial_{\widehat m_\ell}\ln w$.
Combining $Q^2$, $w$, and $d_\ell^{\rm rw}$ on the same configurations
gives the estimator of Eq.~\eqref{eq:staggered-response} at the
simulated quark masses,
\begin{equation}
 \Ctop^{\rm rw}
 =\frac{3}{2}\widehat m_\ell N_4
 \left[
 \frac{\langle Q^2 d_\ell^{\rm rw}\rangle_{\rm rw}}
      {\langle Q^2\rangle_{\rm rw}}
 -\langle d_\ell^{\rm rw}\rangle_{\rm rw}
 \right].
 \label{eq:app-reweighted-dens-response}
\end{equation}

\end{document}